\documentstyle[preprint,aps,prb]{revtex}
\def\dfrac{\displaystyle\frac}
\begin{document}
\draft
\title{Effect of intersubband scattering on weak localization\\
in 2D systems}
\author{N.S.~Averkiev,$^1$ L.E.~Golub,$^{1,2}$  S.A.~Tarasenko\footnote{E-mail: tarasenko@coherent.ioffe.rssi.ru},$^1$ and M.~Willander$^2$} 
\address{$^1$A.F.~Ioffe Physico-Technical Institute, Russian Academy  of Sciences, 194021 St.~Petersburg, Russia}
\address{$^2$Physical Electronics and Photonics, Department of Physics,
Chalmers University of Technology and G\"{o}teborg University, S-412~96, G\"{o}teborg, Sweden}
\date{\today}
\maketitle
\begin{abstract}
The theory of weak localization is generalized for multilevel 2D
systems taking into account intersubband scattering. It is shown that
weak intersubband scattering which is negligible in a classical
transport, affects strongly the weak-localization correction to
conductivity. The anomalous magnetoresistance is calculated
in the whole range of classically low magnetic fields.
This correction to conductivity is shown to depend
strongly on the ratios of occupied level concentrations. It is demonstrated
that at relatively low population of the excited subband,
it is necessary to use the present theory because the high-field limit
asimptotics is shown to be achieved only in classical magnetic fields.
\end{abstract}
\pacs{PACS: 73.20.Fz, 73.63.Hs, 73.21.-b}
\section{Introduction}
Weak localization phenomenon is an interference of waves propagating along
the same paths in opposite directions. Phase and spin relaxation processes
or magnetic field destroy the interference and therefore can make observable the
weak localization effect. The most remarkable manifestation of the
phenomenon is the anomalous behavior of resistance in classically weak
magnetic fields. This effect takes place when the
magnetic length, $l_B$, is simultaneously comparable to characteristic
kinetic lengths.

In very low magnetic fields, when the mean free path, $l$, is much less
than $l_B$ the coherence is lost at long trajectories.
This is so-called {\em diffusion} regime of weak localization, and the
corresponding characteristic size is the dephasing length, $l^\varphi$.
In higher fields, when $l_B \sim l$, the short trajectories passing through
several scatterers contribute to weak localization. This regime is called
{\em non-diffusion}.

The anomalous magnetoresistance was widely investigated both theoretically
and experimentally in bulk semiconductors and metals, thin films and
ultra-quantum two-dimensional (2D) structures. A comparison of theory and
experimental data allowed to
determine the kinetic parameters as times and lengths of elastic relaxation
and dephasing.~\cite{AA}

Recently magnetotransport has been investigated in more
complicated systems which are between 3D and 2D ones. The weak localization
experiments were performed on
2D multivalley semiconductors\cite{Pudalov},
quantum wells with two or several occupied levels of size
quantization,~\cite{Lindelof,Tellur,ZnO}
doped $\delta$-layers,~\cite{Minkov1} and on tunnel-coupled quantum
wells.~\cite{Minkov2} In these quasi-two-dimensional (quasi-2D) systems,
the intersubband scattering takes place. It leads to effective averaging
of the kinetic parameters corresponding to different levels, and
therefore affects a magnitude and behavior of magnetotransport
characteristics.~\cite{ICPS25}
Quasi-2D systems are very attractive objects for study of weak
localization because even rare intersubband transitions affect it strongly.
Usually the scattering time for transition between $\alpha^{th}$
and $\beta^{th}$ subbands, $\tau_{\alpha\beta}$
($\alpha \neq \beta$), exceeds enough the total momentum relaxation
times in subbands, $\tau_{\alpha}$, therefore
an influence of intersubband scattering on classical magnetotransport is
insignificant. In opposite, the dependence of weak-localization correction
to conductivity on magnetic field, $\Delta\sigma(B)$, is determined by
the dephasing time $\tau^{\varphi}_{\alpha}$ exceeding $\tau_{\alpha}$.
Therefore at $\tau_{\alpha\beta}$ long with respect to $\tau_{\alpha}$ but
comparable to $\tau^{\varphi}_{\alpha}$, intersubband scattering affects
weak-localization correction strongly.

To describe the experimental data, one needs the weak-localization theory
taking into account the intersubband scattering for the whole range of
classicaly low magnetic fields. However the theory was developed only for
simple systems with one populated level.\cite{Zyuzin} The intersubband
scattering was considered only in the diffusion
regime.\cite{JapanIntersubband,BergmannDQWS,SSC,FTP,Raichev}

The aim of the present paper is to develop the theory of weak localization
in quasi-2D systems for the whole range of classically low magnetic fields.
It is the theory that is necessary for quantitative determining of kinetic
parameters of structures by comparison with experimental data. 

The paper is organized as follows. In Section~\ref{Theory} we derive the
expressions for the weak-localization correction to the conductivity.
Section~\ref{Discussion} is devoted to the detailed analysis of two
occupied subband system. The main results are given briefly in Conclusion.
                                                                          
\section{Theory}
\label{Theory}

Since intersubband scattering is accompanied by large transfer of the
momentum, it is caused mainly by the short-range part of the potential.
Therefore to attract attention to effect of the intersubband transitions,
the scattering is assumed to be from short-range potential, e.g. impurities.
The main weak-localization corrections to the conductivity appear in the
first order in the parameter $(k_F l)^{-1}$, where $k_F$ is the Fermi
wavevector. The corresponding diagrams for the corrections are presented in
Fig.~\ref{f1}. The dashed lines conform to the correlation function of the total
potential $V(\bbox{\rho},z)$, where $\bbox{\rho}=(x,y)$ are the coordinates
in the 2D plane and $z$ is the perpendicular direction.
Providing $\delta$-scattering, this correlator has the form
\begin{equation}
\left<V(\bbox{\rho},z)V(\bbox{\rho'},z') \right> = W\delta(\bbox{\rho}-\bbox{\rho'})\delta(z-z')f(z) \:.
\end{equation}
Here the averaging is performed over the impurity positions, $f(z)$ is
the function of scatterer distribution, and
$W$ is the factor dependent on intensity and concentration of scatterers.
The solid lines in the diagrams describe the Green functions
of the quasi-2D electron gas in an external magnetic field.
We assume the condition of a ``good conductor'' to be fulfilled for all subbands
\begin{equation}\label{goodconductivity}
\mu_{\alpha} \tau_{\alpha}/\hbar \gg 1 \:,
\end{equation}
and the energy distances between subbands are large
\begin{equation}
\label{largedelta}
|\mu_{\alpha} - \mu_{\beta}| \gg \hbar / \tau_{\alpha} \:.
\end{equation}
Here $\mu_{\alpha}$ is the chemical potential counted from the bottom of the
$\alpha^{th}$ subband,  and $\tau_{\alpha}$ is the momentum relaxation time
of carriers in the subband,
\begin{equation}
1/\tau_{\alpha}=\sum_{\beta}1/\tau_{\alpha\beta} \:,
\end{equation}
where $\tau_{\alpha\beta}=\tau_{\beta\alpha}$ is the scattering time between
the $\alpha^{th}$ and $\beta^{th}$ subbands, and summation is performed
over all occupied levels of size quantization. Since $\tau_{\alpha\beta}$
arises due to scattering from impurities, it is defined by the following
relation
\begin{equation}
1/\tau_{\alpha\beta}=\frac{2\pi}{\hbar}N_F W \int f(z) u^2_{\alpha}(z)
u^2_{\beta}(z) dz \:,
\end{equation}
where $N_F=m/(2\pi\hbar^2)$ is the density of states at Fermi level
with a fixed spin, and $m$ is the electron effective mass in the 2D plane.

The advanced and retarded Green functions are diagonal in subband
indices under conditions~(\ref{goodconductivity}),~(\ref{largedelta})~\cite{Raikh,SdH}
\begin{eqnarray}\label{GRaikh}
G^{A,R}(\bbox{\rho},z,\bbox{\rho'},z')=\sum_{\alpha}
G^{A,R}_{\alpha}(\bbox{\rho},\bbox{\rho'}) u_{\alpha}(z)u_{\alpha}(z') \:,\nonumber \\
G^{A,R}_{\alpha}(\bbox{\rho},\bbox{\rho'})= \sum_{Nk_y}
\frac{ \psi_{N k_y}(\bbox{\rho}) \psi^{*}_{N k_y}(\bbox{\rho'})}
{\mu_{\alpha}-\hbar \omega_c (N+1/2) \pm i\hbar/2\tau_{\alpha}\pm i\hbar/2\tau_{\alpha}^{\varphi}} \:.
\end{eqnarray}
Here $\psi_{N k_y}(\bbox{\rho})u_{\alpha}(z)$ is the electron wave function
in the heterostructure in the Landau gauge with the vector potential
directed along the $y$ axis,
$N$ and $k_y$ are the number of the Landau level and the value of the
in-plane wavevector, $\omega_c$ is
the cyclotron frequency, and the dephasing time $\tau_{\alpha}^{\varphi}$
($\tau_{\alpha}^{\varphi}\gg \tau_{\alpha}$) describes inelastic scattering. 

In a classically weak magnetic field, when $\omega_c\tau_{\alpha} \ll 1$, the field dependence appears only in the phase of the Green functions, similarly to the case of one subband occupation:
\begin{equation}\label{G}
G_{\alpha}^{A,R}(\bbox{\rho},\bbox{\rho'})=
\exp\left[i\Phi(\bbox{\rho},\bbox{\rho'})\right]
G_{\alpha}^{(0)A,R}(\bbox{\rho}-\bbox{\rho'}) \:.
\end{equation}
The phase factor is given by
\begin{equation}
\Phi(\bbox{\rho},\bbox{\rho'})=\frac{(y'-y)(x'+x)}{2 l^2_B} \:,
\end{equation}
and $G^{(0)A,R}$ are the Green functions in zero mangetic field
\begin{equation}\label{G0AR}
G_{\alpha}^{(0)A,R}(\bbox{\rho}-\bbox{\rho'})=
- 2 N_F K_{0}\left[\pm i k_{\alpha}|\bbox{\rho}-\bbox{\rho'}|+
\frac{|\bbox{\rho}-\bbox{\rho'}|}{2l_{\alpha}}
\left(1+\frac{l_{\alpha}}{l^{\varphi}_{\alpha}} \right) \right] \:,
\end{equation}
where $l_B=\sqrt{\hbar c/|e|B}$ is the magnetic length, with $e<0$ and
$B$ being the electron charge and magnetic field, $K_0$ is the McDonald function,
$k_{\alpha}$, $l_{\alpha}=\hbar k_{\alpha} \tau_{\alpha} /m$ and
$l_{\alpha}^{\varphi}=\hbar k_{\alpha} \tau^{\varphi}_{\alpha} /m$
are the Fermi wavevector, the mean free path and the dephasing length
in the $\alpha^{th}$ subband, respectively.
  
Figs.~\ref{f1}a-d show all types of diagrams contributing to the
anomalous magnetoconductivity. The similar diagrams with two scattering
lines are depicted in Fig.~\ref{f1}e.

The latter are special. In the papers~\cite{EbisawaFukuyama,Kawabata} they were
calculated incorrect way and therefore a non-physical divergence was
obtained. It was noted in Ref.~\onlinecite{Zyuzin} that two-impurity line
diagrams  correspond to non-self-crossing trajectories which have zero magnetic
flux passing through them and, hence, a field does not change this
correction to the conductivity, $\Delta \sigma_2$.
In Ref.~\onlinecite{France} the contribution was investigated in more
detail. It was shown formally that $\Delta \sigma_2$ is independent of
magnetic field and its value has an order of $e^2/\hbar$. However the
calculation of $\Delta \sigma_2$ was not performed. It was claimed in
Ref.~\onlinecite{Dmitriev} that $\Delta \sigma_2 =0 $,
however the proof of this statement was absent.

Thus, $\Delta \sigma_2$ i) has never been calculated and ii) it was
claimed without proof that $\Delta \sigma_2 = 0$. Below we calculate
this contribution and show that it is not equal to zero.

\subsection{Contribution from the diagrams with two scattering lines}

The expression for the conductivity correction from
the diagrams with two scattering lines (Fig.~\ref{f1}e) has the form
\begin{eqnarray}
\label{dsigma2}
\Delta\sigma_2=\frac{\hbar}{2\pi} \sum_{\alpha\beta\gamma\delta}
W^2_{\alpha\beta\gamma\delta} \int \int d \bbox{\rho_1} d \bbox{\rho_2} \,
\: \bbox{J}^{RA}_{\alpha}(\bbox{\rho_2},\bbox{\rho_1}) \cdot
\left[
\bbox{J}^{AR}_{\beta}(\bbox{\rho_2},\bbox{\rho_1})
G^{A}_{\gamma}(\bbox{\rho_1},\bbox{\rho_2})
G^{R}_{\delta}(\bbox{\rho_1},\bbox{\rho_2}) +  \right. \nonumber \\
\left.
\bbox{J}^{AR}_{\beta}(\bbox{\rho_1},\bbox{\rho_2})
G^{A}_{\gamma}(\bbox{\rho_1},\bbox{\rho_2})
G^{A}_{\delta}(\bbox{\rho_2},\bbox{\rho_1}) +
\bbox{J}^{AR}_{\beta}(\bbox{\rho_1},\bbox{\rho_2})
G^{R}_{\gamma}(\bbox{\rho_1},\bbox{\rho_2})
G^{R}_{\delta}(\bbox{\rho_2},\bbox{\rho_1}) 
\right] \:.
\end{eqnarray}
Here the spin degeneracy has been taken into account,
\begin{equation}
W_{\alpha\beta\gamma\delta}=W\int f(z) u_{\alpha}(z) u_{\beta}(z) u_{\gamma}(z)  u_{\delta}(z) dz \:,
\end{equation} 
and the current vertex is defined as 
\begin{equation}\label{vertexdef}
\bbox{J}^{RA}_{\alpha}(\bbox{\rho},\bbox{\rho'})=
\int d  \bbox{\rho_1} G^{R}_{\alpha}(\bbox{\rho},\bbox{\rho_1}) \: \hat{\bbox J} (\bbox{\rho_1}) \:G^{A}_{\alpha}(\bbox{\rho_1},\bbox{\rho'}) 
\:,
\end{equation}
where $\hat{\bbox J}$ is the electric current operator. In classically low
magnetic fields, the vertices have the form
\begin{equation}\label{vertexB}
\bbox{J}^{RA}_{\alpha}(\bbox{\rho},\bbox{\rho'})=
\bbox{J}^{AR}_{\alpha}(\bbox{\rho},\bbox{\rho'})=
\exp\left[i\Phi(\bbox{\rho},\bbox{\rho'})\right]
\bbox{J}^{(0)}_{\alpha}(\bbox{\rho}-\bbox{\rho'}) \:,
\end{equation}
where $\bbox{J}^{(0)}$ is the current vertex in zero magnetic field
\begin{equation}\label{vertex0}
\bbox{J}^{(0)}_{\alpha}(\bbox{\rho}-\bbox{\rho'})=
\frac{e\tau_{\alpha}}{m} \bbox{\nabla_{\rho}}
\left[ G^{(0)R}_{\alpha}(\bbox{\rho}-\bbox{\rho'})  -
G^{(0)A}_{\alpha}(\bbox{\rho}-\bbox{\rho'}) \right] \:.
\end{equation}

One can see, the conductivity correction $\Delta\sigma_2$ is independent of magnetic
field, because the current vertices in~(\ref{dsigma2}),
$\bbox{J}_{\alpha}(\bbox{\rho_2},\bbox{\rho_1})$, contain the phase
factor which is complex conjugated to that contained in
$G_{\beta}(\bbox{\rho_1},\bbox{\rho_2})$.

Taking into account~(\ref{G}),~(\ref{G0AR}),~(\ref{vertexB}),~(\ref{vertex0}) and 
providing the limits $k_{\alpha}l_{\alpha}\rightarrow\infty$, 
$l_{\alpha}/ l_{\alpha}^\varphi \rightarrow\infty$, we obtain  the contribution of the diagrams with two impurity lines
\begin{eqnarray}
\label{dsigma2K0K1}
\Delta\sigma_2 &=& - \frac{e^2}{\pi^4 \hbar} \sum_{\alpha\beta\gamma\delta}
\left(\frac{2\pi}{\hbar}N_F W_{\alpha\beta\gamma\delta} \right)^2\,\tau_{\alpha}\tau_{\beta}
\nonumber \\
&& \times  k_\alpha k_\beta \int\limits_{0}^{\infty}d \rho \rho
\left[ K_1\left(ik_{\alpha}\rho\right) +
K_1\left(-ik_{\alpha}\rho\right)
\right] 
\left[ K_1\left(ik_{\beta}\rho\right) +
K_1\left(-ik_{\beta}\rho\right)
\right]
\nonumber \\
&& \times \left[ K_0\left(ik_{\gamma}\rho\right) K_0\left(-ik_{\delta}\rho\right) \right.
- \left. K_0\left(ik_{\gamma}\rho\right) K_0\left(ik_{\delta}\rho\right) -
K_0\left(-ik_{\gamma}\rho\right)
K_0\left(-ik_{\delta}\rho\right)
\right]  \:.
\end{eqnarray}
The asymptotic expansions for $K_0$ and $K_1$ at large arguments have
not to be used for calculation of  $\Delta\sigma_2$, otherwise
the integrand tends to infinity at $\rho \rightarrow 0$. It is the
mistake that was done in Refs.~\onlinecite{EbisawaFukuyama,Kawabata}.

Eq.~(\ref{dsigma2K0K1}) shows that the diagrams with two impurity lines give
rise to non-zero contribution to the conductivity contrary to the notice
in Ref.~\onlinecite{Dmitriev}. 
For the case of one occupied subband, we obtain (see Appendix)
\begin{equation}
\Delta\sigma_2 =-\frac{e^2}{\pi^2 \hbar} \ln{2} \:.
\end{equation}
                            
\subsection{Magnetoconductivity calculation}

Now, we consider the weak-localization corrections which do contribute
to the anomalous magnetoresistance. The corresponding diagrams are presented
in Fig.~\ref{f1}a-d. One can show the terms c) and d) cancel out each other
similarly to the case of one occupied subband. Thus the field-sensitive
conductivity correction has the form
\begin{equation}
\Delta\sigma = \Delta\sigma^{(a)} + \Delta\sigma^{(b)}\:,
\end{equation}
\begin{eqnarray}
&\mbox{}&\Delta\sigma^{(a)}=\frac{\hbar}{2\pi} \sum_{\alpha}
\int d \bbox{\rho_1} d \bbox{\rho_2} \:
\bbox{J}^{RA}_{\alpha}(\bbox{\rho_2},\bbox{\rho_1}) \cdot
\bbox{J}^{AR}_{\alpha}(\bbox{\rho_2},\bbox{\rho_1}) \:
{\cal C}^{(3)}_{\alpha\alpha}(\bbox{\rho_1},\bbox{\rho_2}) \:, \\
\Delta\sigma^{(b)}&=&\frac{\hbar}{\pi} \sum_{\alpha\beta}
\int d \bbox{\rho_1} d \bbox{\rho_2} d \bbox{\rho_3}
\left[
J^{RA}_{x\alpha}(\bbox{\rho_3},\bbox{\rho_1})
J^{AR}_{x\beta}(\bbox{\rho_1},\bbox{\rho_2})
{\cal C}^{(2)}_{\beta\alpha}(\bbox{\rho_2},\bbox{\rho_3})
G^{A}_{\beta}(\bbox{\rho_1},\bbox{\rho_2})
G^{A}_{\alpha}(\bbox{\rho_3},\bbox{\rho_1}) \right. \nonumber \\
&+&\left. 
J^{RA}_{x\alpha}(\bbox{\rho_1},\bbox{\rho_2}) 
J^{AR}_{x\beta}(\bbox{\rho_3},\bbox{\rho_1})
{\cal C}^{(2)}_{\alpha\beta}(\bbox{\rho_2},\bbox{\rho_3})
G^{R}_{\beta}(\bbox{\rho_3},\bbox{\rho_1})
G^{R}_{\alpha}(\bbox{\rho_1},\bbox{\rho_2}) \right]
W_{\alpha\alpha\beta\beta} \:,
\end{eqnarray}
where $J_{x\alpha}$ is the $x$-projection of the vector $\bbox{J}_{\alpha}$,
and the Cooperons ${\cal C}^{(2)}$ and ${\cal C}^{(3)}$ are the sums of the fan
internal parts of the corresponding diagrams starting with two and three lines, respectively. In general, these
parts depend on four subband indices. However one can show the Cooperons
are diagonal in the pairs of indices due to the
relations~(\ref{goodconductivity}),~(\ref{largedelta}) similarly to the
Green functions (see Eq.~\ref{GRaikh}).

The Cooperons ${\cal C}^{(2)}$ and ${\cal C}^{(3)}$ can be found from the
following systems of equations
\begin{eqnarray}\label{systemC2C3}
{\cal C}^{(2)}_{\alpha\beta}(\bbox{\rho},\bbox{\rho'}) &=&
{2 \pi N_F \over \hbar} \sum_{\gamma}W_{\alpha\alpha\gamma\gamma}W_{\gamma\gamma\beta\beta} \tau_\gamma
P_{\gamma}(\bbox{\rho},\bbox{\rho'}) \nonumber \\
&+& {2 \pi N_F \over \hbar} \sum_{\gamma}W_{\alpha\alpha\gamma\gamma}
\tau_\gamma \int d \bbox{\rho_1}
P_{\gamma}(\bbox{\rho},\bbox{\rho_1})
{\cal C}^{(2)}_{\gamma\beta}(\bbox{\rho_1},\bbox{\rho'}) \:,\\
{\cal C}^{(3)}_{\alpha\beta}(\bbox{\rho},\bbox{\rho'})&=&
{\cal C}^{(2)}_{\alpha\beta}(\bbox{\rho},\bbox{\rho'})-
{2 \pi N_F \over \hbar} \sum_{\gamma}W_{\alpha\alpha\gamma\gamma}W_{\gamma\gamma\beta\beta}
\tau_\gamma P_{\gamma}(\bbox{\rho},\bbox{\rho'}) \nonumber \:,
\end{eqnarray}
where 
\begin{equation}
\label{defP}
P_{\alpha}(\bbox{\rho},\bbox{\rho_1})=
{\hbar \over 2 \pi N_F \tau_\alpha} G^{A}_{\alpha}(\bbox{\rho},\bbox{\rho_1})
G^{R}_{\alpha}(\bbox{\rho},\bbox{\rho_1}) \:.
\end{equation}

In order to solve the systems~(\ref{systemC2C3}), we expand
the kernel of the integral operator, $P_{\alpha}(\bbox{\rho},\bbox{\rho_1})$, in series of 
the 2D wave functions of a particle with the mass $m$ and the charge $2e$
under a perpendicular magnetic field in the Landau gauge,
$\chi_{N k_y}(\bbox{\rho})$. In this basis, $P_{\alpha}(\bbox{\rho},\bbox{\rho_1})$ is diagonal:~\cite{Zyuzin,Kawabata}
\begin{equation}
\label{hiexpansionP}
P_{\alpha}(\bbox{\rho},\bbox{\rho_1})=
\sum_{N k_y} P_{\alpha}(N)
\chi_{N k_y}(\bbox{\rho}) \chi^*_{N k_y}(\bbox{\rho_1}) \:.
\end{equation}
From Eqs.~(\ref{systemC2C3}) it follows that ${\cal C}^{(2)}_{\alpha\beta}(\bbox{\rho},\bbox{\rho_1})$ and
${\cal C}^{(3)}_{\alpha\beta}(\bbox{\rho},\bbox{\rho_1})$ are also diagonal in this basis:
\begin{equation}\label{hiexpansionC}
{\cal C}^{(2,3)}_{\alpha\beta}(\bbox{\rho},\bbox{\rho_1})=\frac{\hbar}{2\pi N_F}
\sum_{N k_y} {\cal C}^{(2,3)}_{\alpha\beta}(N)
\chi_{N k_y}(\bbox{\rho}) \chi^*_{N k_y}(\bbox{\rho_1}) \:.
\end{equation}

The asymptoticses for the Green functions and the current
vertex at long distances, $|\bbox{\rho}-\bbox{\rho'}|k_{\alpha} \gg 1$,
can be used for calculation of the diagrams with three or more dashed lines.
The corresponding expressions are
\begin{equation}\label{G0approx}
G_{\alpha}^{(0)A,R}(\bbox{\rho}-\bbox{\rho'}) \approx -N_F
\sqrt{\frac{2\pi}{k_{\alpha}|\bbox{\rho}-\bbox{\rho'}|}}
\exp\left[ \mp i \frac{\pi}{4} \mp i k_{\alpha}|\bbox{\rho}-\bbox{\rho'}|
-\frac{|\bbox{\rho}-\bbox{\rho'}|}{2l_{\alpha}}
\left(1+\frac{l_{\alpha}}{l^{\varphi}_{\alpha}}\right) \right] \:,
\end{equation}
\begin{equation}\label{J0approx}
\bbox{J}^{(0)}_{\alpha}(\bbox{\rho}-\bbox{\rho'}) \approx
i\frac{e}{\hbar} l_{\alpha}
\left[ G^{(0)R}_{\alpha}(\bbox{\rho}-\bbox{\rho'}) +
G^{(0)A}_{\alpha}(\bbox{\rho}-\bbox{\rho'}) \right]
\frac{\bbox{\rho}-\bbox{\rho'}}{|\bbox{\rho}-\bbox{\rho'}|} \:.
\end{equation}
Thus, the coefficients for the expansion of the kernel are given by
\begin{equation}
P_{\alpha}(N)=\frac{l_B}{l_{\alpha}}\int\limits_{0}^{\infty} dx
\exp\left[ -x\frac{l_B}{l_{\alpha}}
\left(1+\frac{l_{\alpha}}{l^{\varphi}_{\alpha}} \right)
-\frac{x^2}{2} \right] L_N(x^2) \:,
\end{equation}
where $L_N$ are the Laguerre polinomials. The values ${\cal C}^{(2)}(N)$
and  ${\cal C}^{(3)}(N)$ are defined by the following systems of linear equations
\begin{eqnarray}
\label{systemC}
\sum_{\gamma}\left( \delta_{\alpha\gamma} -
\frac{\tau_{\gamma}}{\tau_{\alpha\gamma}} P_{\gamma}(N) \right)
{\cal C}^{(2)}_{\gamma\beta}(N) &=& \sum_{\gamma}
\frac{\tau_{\gamma}}{\tau_{\alpha\gamma}\tau_{\gamma\beta}}
P_{\gamma}(N) \:,\nonumber \\
{\cal C}^{(3)}_{\alpha\beta}(N)={\cal C}^{(2)}_{\alpha\beta}(N)
&-& \sum_{\gamma}
\frac{\tau_{\gamma}}{\tau_{\alpha\gamma}\tau_{\gamma\beta}}
P_{\gamma}(N) \:.
\end{eqnarray}
Neglecting the rapidly oscillating terms $G^{R}G^{R}$ and $G^{A}G^{A}$ and taking
into accout the expansions~(\ref{hiexpansionP}),~(\ref{hiexpansionC}), we obtain
\begin{equation}\label{dsigmaa}
\Delta\sigma^{(a)}=-\frac{e^2}{\pi^2\hbar} \sum_{\alpha}
\frac{l_{\alpha}^2}{l_B^2} \tau_{\alpha}
\sum_{N=0}^{\infty}P_{\alpha}(N)\,{\cal C}^{(3)}_{\alpha\alpha}(N) \:.
\end{equation}
Expanding the expression $P_{\alpha}(\bbox{\rho},\bbox{\rho'})
\dfrac{\bbox{\rho}-\bbox{\rho'}}{|\bbox{\rho}-\bbox{\rho'}|}$
in series of the functions $\chi_{N k_y}(\bbox{\rho})$, one can also find
\begin{equation}\label{dsigmab}
\Delta\sigma^{(b)}=\frac{e^2}{\pi^2\hbar} \sum_{\alpha\beta}
\frac{l_{\alpha}l_{\beta}}{l_B^2}
\frac{\tau_{\alpha}\tau_{\beta}}{\tau_{\alpha\beta}}
\sum_{N=0}^{\infty}Q_{\alpha}(N)Q_{\beta}(N)
\frac12\left[{\cal C}^{(2)}_{\alpha\beta}(N)+{\cal C}^{(2)}_{\beta\alpha}(N+1)
\right]\:,
\end{equation}
where
\begin{equation}
Q_{\alpha}(N)=\frac{l_B}{l_{\alpha}}\frac{1}{\sqrt{N+1}}
\int\limits_{0}^{\infty} dx \: x
\exp\left[ -x\frac{l_B}{l_{\alpha}}
\left(1+\frac{l_{\alpha}}{l^{\varphi}_{\alpha}} \right)
-\frac{x^2}{2} \right] L_N^1(x^2) \:.
\end{equation}
with $L_N^1$ being the associated Laguerre polinomials.

Eqs.~(\ref{dsigmaa},\ref{dsigmab}) describe the weak-localization
correction to conductivity, $\Delta\sigma$,
in the whole range of classically-weak fields, $\omega_c \tau \ll 1$,
when $l_B$ may be both larger and smaller than $l_\alpha$. Now we
consider the limiting cases.

In the zero-field limit, the large number of terms, up to $N \sim (l_B/l_\alpha)^2$,
is essential in the sums~(\ref{dsigmaa},\ref{dsigmab}).
Therefore to provide the calculation of $\Delta\sigma(0)$,
the summation over $N$ should be replaced by integration with the
following zero-field asymptoticses
\begin{displaymath}
P_{\alpha}(N) \approx  \frac{1}{\sqrt{(1+l_{\alpha}/l^{\varphi}_{\alpha})^2
+4N(l_{\alpha}/l_B)^2}} \:, \;\;\;
Q_{\alpha}(N) \approx \frac{1-(1+
l_{\alpha}/l^{\varphi}_{\alpha})P_{\alpha}(N)}
{2\sqrt{N}l_{\alpha}/l_B} \:.
\end{displaymath}

In low fields,
\begin{equation}
\label{low_fields}
l_B \sim \sqrt{l_\alpha l_\alpha^{\varphi}} \gg l_\alpha \:,
\end{equation}
the so-called diffusion approximation is valid.
The interference is destroyed at long trajectories, where particles
experience many scattering events and, hence, their motion is a diffusion.
In the frame of the approach, one can calculate the difference between
quantum corrections in finite and zero fields
$\Delta\sigma (B) - \Delta\sigma(0)$. To obtain the expression, the
Cooperons determined from the system~(\ref{systemC}) should be sought
in the form of diffusion poles.~\cite{FTP}
The difference $\Delta\sigma (B) - \Delta\sigma(0)$ due to the diagrams~b) in Fig.~\ref{f1} are small in fields~(\ref{low_fields}), 
and therefore the correction behavior is defined by the low-field asymptotics
of $\Delta\sigma^{(a)}$ only. The value of the conductivity correction
in zero field itself, $\Delta\sigma(0)$, calculated with the diffusion approach
is correct at $\ln{(\tau_\alpha^\varphi/\tau_\alpha)} \gg 1$ that
is not realized practically.

In the particular case of two occupied subband system,
the diffusion approximation yields the following
magnetoconductivity correction
\begin{equation}
\label{dsigma_diff}
\Delta\sigma (B) - \Delta\sigma(0) =
{e^2 \over 2 \pi^2 \hbar}  \left[
f_2 \left( \frac{\tilde{l}_1^2}{l_B^2}\right) +
f_2 \left( \frac{\tilde{l}_2^2}{l_B^2}\right)  \right]  \:,
\end{equation}
where $f_2(x) = \ln x + \psi (1/2 + 1/x)$, and $\psi (y)$ is the
digamma-function. The lengths $\tilde{l}_{1,2}$ are given by~\cite{SSC,FTP}
$$
\tilde{l}_{1,2}^2 =
\dfrac{4 \: l_1 l_2}{\dfrac{l_2}{l_1}\dfrac{\tau_1}{t_1} +
\dfrac{l_1}{l_2} \dfrac{\tau_2}{t_2} \pm
\sqrt{ \left( \dfrac{l_2}{l_1}\dfrac{\tau_1}{t_1} -
\dfrac{l_1}{l_2} \dfrac{\tau_2}{t_{2}} \right)^2 +
4 \dfrac{\tau_1 \tau_2}{\tau_{12}^2}}} \:,
$$
where the corresponding times are
$$
\dfrac{1}{t_{1,2}} = \dfrac{1}{\tau_\varphi^{(1,2)}} +
\dfrac{1}{\tau_{12}} \:.
$$

In the high-field limit, $l_B \ll l_{\alpha}$, the weak-localization corrections
to the conductivity have the following asymptoticses
\begin{eqnarray}\label{highB}
\Delta\sigma^{(a)}&\approx& -\frac{e^2}{\pi^2\hbar} \lambda^{(a)}
\sum_{\alpha\beta\gamma} \frac{l_B l_{\alpha}}{l_{\beta}l_{\gamma}}
\frac{\tau_{\alpha}\tau_{\beta}\tau_{\gamma}}
{\tau_{\alpha\beta}\tau_{\beta\gamma}\tau_{\gamma\alpha}} \:, \nonumber
\\
\Delta\sigma^{(b)}&\approx& \frac{e^2}{\pi^2\hbar} \lambda^{(b)}
\sum_{\alpha\beta\gamma} \frac{l_B}{l_{\alpha}}
\frac{\tau_{\alpha}\tau_{\beta}\tau_{\gamma}}
{\tau_{\alpha\beta}\tau_{\beta\gamma}\tau_{\gamma\alpha}} \:, 
\end{eqnarray}
where the constants $\lambda^{(a)}$ and $\lambda^{(b)}$ are given by 
\begin{eqnarray}
\lambda^{(a)}&=&\sum\limits_{N=0}^{\infty}\left[\int\limits_{0}^{\infty}
dx \exp{\left(-{x^2 \over 2} \right)} L_N(x^2) \right]^3 \approx 2.7 \:,
\\
\lambda^{(b)}&=&\sum\limits_{N=0}^{\infty}
\frac{1}{N+1}\left[\int\limits_{0}^{\infty} dx x \exp{\left(-{x^2 \over 2} \right)} L^1_N(x^2)
\right]^2 \left[\int\limits_{0}^{\infty} dx \exp{\left(-{x^2 \over 2} \right)}
\frac{L_N(x^2) + L_{N+1}(x^2)}{2} \right] \approx 0.94 \:. \nonumber
\end{eqnarray}

\section{Results and discussion}
\label{Discussion}

In this section, we consider the system with two size-quantized levels
in detail. To make the consideration closer to experimental applications,
each subband will be described with the concentration of carriers
$n_{\alpha} = k^2_{\alpha}/2\pi$.

In Fig.~\ref{f2} the solid curves present the dependence of weak-localization
correction to the conductivity, $\Delta\sigma$, on magnetic field at
various intersubband scattering rates and different level
occupations. We assume here that the total relaxation
times in the subbands coincide, $\tau_1=\tau_2$, and the dephasing
times are also identical and equal to $10\tau_1$. Fig.~\ref{f2}a shows the
case of frequent intersubband transitions,
$\tau_{12} \sim \tau_1 \ll \tau_1^{\varphi}$. Fig.~\ref{f2}b corresponds to
relatively rare transitions when the intersubband scattering time is
comparable to that of dephasing,
$\tau_{12} \sim \tau_1^{\varphi} \gg \tau_1$. Fig.~\ref{f2}c depicts the
case of isolated levels,
$\tau_{12} \gg \tau_1,\tau_1^{\varphi}$.

In the presence of intensive intersubband scattering (Fig.~\ref{f2}a), the
magnitude of the weak-localization conductivity correction
depends strongly on the
excited level occupation in the whole range of magnetic fields.
In the opposite case of the relatively weak intersubband scattering,
when $\tau_{12} \gg \tau_1$ (Figs.~\ref{f2}b,~\ref{f2}c), $\Delta\sigma$ is practically
independent of $n_2/n_1$ in zero field. The dependence appears in
finite fields only, when the magnetic
length is comparable with the mean free paths. The reason is the
influence of subband concentration on the mean free path which determines
the scale of the changing of the weak-localization correction in magnetic
field but the value of $\Delta\sigma$ at $B=0$.

If intersubband scattering
is weak enough (Figs.~\ref{f2}b,\ref{f2}c), its role is restricted
to additional dephasing. Therefore the difference between the
curves presented in Figs.~\ref{f2}b,\ref{f2}c and corresponded to the
same ratio $n_2/n_1$ appears in low fields rather.

The changes of the weak-localization correction in magnetic
field, $\Delta\sigma(B)-\Delta\sigma(0)$, calculated in the frame of
diffusion approximation with Eq.~(\ref{dsigma_diff}) are presented in
Fig.~\ref{f2} with the dashed curves.
This theory is seen to give the reduced absolute value of $\Delta\sigma$
in intermediate fields so that $l_B < \sqrt{l l^{\varphi}}$
(see Eq.~(\ref{low_fields})).

The dotted curves in Fig.~\ref{f2} present the high-field limit
($l_B \ll l_1,l_2$) dependences of the anomalous
magnetoconductivity~(\ref{highB}). If the subband concentrations
are comparable ($n_2/n_1=0.5$ in Fig.~\ref{f2}), $l_1 \sim l_2$, then
the weak-localization correction reaches its asymptotic behavior
in the range of classically weak magnetic fields. In the opposite case,
$l_2 \ll l_1$, the asymptotics of $\Delta\sigma$ corresponding to the
excited subband takes place at so high fields that the
magnetoconductivity of the ground subband is likely to be of classical
nature.

Thus, Fig.~\ref{f2} shows clearly that the anomalous magneticonductivity in the
whole magnetic field range should be described with the exact
expressions~(\ref{dsigmaa},\ref{dsigmab})
particularly in the case of relatively small excited subband occupation.

Fig.~\ref{f3} presents the dependence of the quantum conductivity correction in
zero magnetic field, $\Delta\sigma(0)$, on the ratio of the subband
concentrations. In the absence of intersubband scattering (solid curve),
$\Delta\sigma(0)$ is not affected by the ratio $n_2/n_1$ because
the correction $\Delta\sigma^{(a)}(0)+\Delta\sigma^{(b)}(0)$
of any independent 2D level is universal and equals to
$-\dfrac{e^2}{2\pi^2\hbar}\ln{\left(\dfrac{\tau^{\varphi}}{2\tau}\right)}$.
If intersubband transitions take place then $\Delta\sigma(0)$ does
depend on $n_2/n_1$. However at weak scattering, $\tau_{12} \gg \tau_1$
(dashed curve), the dependence is insignificant. The effect of intersubband
scattering is the decreasing of the absolute value, $|\Delta\sigma(0)|$,
with respect to the isolated level case. Since weak intersubband scattering
acts as an additional dephasing, the effective dephasing time becomes
shorter and therefore $|\Delta\sigma(0)|$ decreases. At intensive scattering,
$\tau_{12} \sim \tau_1$ (dotted curve), the magnitude of the conductivity
correction changes in several times in the shown range of the
concentration ratio.

Moreover, one can say that the increasing of the intersubband scattering
intensity causes the transition from two-level into one-level system.
Indeed, in the absence of intersubband scattering, two independent levels
exist. In the case of intensive intersubband scattering,
$\tau_{12} \sim \tau_1$, the level division does not take place. There is
only one subband effectively with the averaged kinetic parameters.
Since the total and dephasing subband times are chosen to be identical
respectively, $\tau_1=\tau_2$, $\tau^{\varphi}_1=\tau^{\varphi}_2$,
the average parameters of the 'effective subband' coincide with those of
separate subbands at the same level occupations, $n_1=n_2$.
The one-level weak-localization correction to conductivity is
universal and independent of the
level occupation. Therefore at $n_1=n_2$ the magnitude of
$|\Delta\sigma|$ for intensive intersubband scattering is half as much
as for isolated subband system. This difference by a factor of 2 in zero field
can be seen in Fig.~\ref{f3}. In the case of the arbitrary level
concentration ratio the quantum conductivity correction depends on the
intersubband scattering intensity in complicated manner.

Fig.~\ref{f4} demonstrates the high-field limit asymptotics of the conductivity
correction~(\ref{highB}), $-\Delta\sigma(B) \times l_1/l_B$, as a function of level occupation ratio.
The analysis of
Eq.~(\ref{highB}) shows the averaging caused by intersubband scattering
is complicated and even can change the functional dependence of
$\Delta\sigma$ on the level concentrations. It is demonstrated by
the curves presented in Fig.~\ref{f4} and corresponded to different intersubband
scatering rates.
Specifically, in the absence of intersubband scattering (solid curve)
the levels contribute to the conductivity correction independently
\begin{equation}\label{highBweak}
\Delta\sigma(B)\approx -\frac{e^2}{\pi^2\hbar}
(\lambda^{(a)}-\lambda^{(b)})
\left( \frac{l_B}{l_1} + \frac{l_B}{l_2} \right) \:.
\end{equation}
In the case of very intensive intersubband scattering (dotted curve)
the high-field limit expression has the form
\begin{equation}\label{highBint}
\Delta\sigma(B)\approx -\frac{e^2}{\pi^2\hbar} \frac12
\left[ \lambda^{(a)} \frac{l_B(l_1+l_2)^3}{4l^2_1 l^2_2}
- \lambda^{(b)} \left( \frac{l_B}{l_1} + \frac{l_B}{l_2} \right)
\right] \:.
\end{equation}
One can see the quantum corrections obtained with~(\ref{highBweak})
and~(\ref{highBint}) at the same level concentrations, $n_1=n_2$,
differs by a factor of 2 similarly to the zero field case.

The dependences described in Figs.~\ref{f2}-\ref{f4} are able to be traced in real
quasi-2D structures by variation of the excited subband population
as it has been performed in other magnetotransport experiments
(see e.g. Ref.~\onlinecite{Zaremba}).
\section{Conclusion}
The theory of weak localization has been developed for multilevel 2D systems
in the whole range of classically weak magnetic fields. Both
the diffusion approximation and high-field limit asymptotics have turned
out to describe the magnetoconductivity behavior in very
narrow field ranges. For the first time the contribution to the
conductivity from all of the self-crossed diagrams has been calculated.
It has been shown that the weak-localization
correction to conductivity depends strongly on the level concentration
ratio and intersubband scattering intensity. Specifically, at the
comparable level occupations and the same relaxation times,
the conductivity correction of the $M$-subband system
decreases in $M$ times when transiting from the isolated
levels to the case of intensive intersubband scattering.
The detailed calculations have been performed for the widely investigated
system with two occupied size-quantized levels. The results are
presented in the form making allowance for comparison
with experimental observations.
                                    
\section*{Acknowledgements}
This work was supported by the Russian Foundation for Basic Research, projects
00-02-17011 and 00-02-16894, and by the Russian State Programme ``Physics of Solid State
Nanostructures''.

\begin{appendix}
\section{}
Let introduce the Fourier-images of the Green fuctions in zero magnetic
field~(\ref{G0AR})
\begin{equation}
\label{Gk}
G^{A,R}_{\alpha}(\bbox{k})=
\frac{1}{\mu_{\alpha}-\hbar^2 k^2 / 2 m \pm i\hbar/2\tau_{\alpha}\pm i\hbar/2\tau_{\alpha}^{\varphi}} \:.
\end{equation}
From~(\ref{vertex0}) it follows that
\begin{equation}
\label{Jk}
\bbox{J}^{(0)}_{\alpha}(\bbox{\rho}-\bbox{\rho'})=
\frac{i e \tau_{\alpha}}{m} \sum_{\bf k} \bbox{k}
\left[ G^{R}_{\alpha}(\bbox{k})  -
G^{A}_{\alpha}(\bbox{k}) \right] \exp{[i \bbox{k}
\cdot (\bbox{\rho}-\bbox{\rho'})]}\:.
\end{equation}
Using Eqs.~(\ref{dsigma2},\ref{Gk},\ref{Jk}) we obtain the following expression for $\Delta\sigma_2$
\begin{eqnarray}
\label{ds2k}
\Delta\sigma_2=
- \frac{e^2 \hbar}{2\pi m^2} \sum_{\alpha\beta\gamma\delta}
W^2_{\alpha\beta\gamma\delta} \tau_\alpha \tau_\beta
\sum\limits_{\bf{k_1} \bf{k_2} \bf{k_3}} \bbox{k_1} \cdot \bbox{k_2}
\left[ G^{A}_{\alpha}(\bbox{k_1}) - G^{R}_{\alpha}(\bbox{k_1}) \right] 
\left[ G^{A}_{\beta}(\bbox{k_2}) - G^{R}_{\beta}(\bbox{k_2}) \right] \\
\left[ 
G^{A}_{\gamma}(\bbox{k_3}) G^{R}_{\delta}(\bbox{k_1}+\bbox{k_2}+\bbox{k_3}) 
+ G^{A}_{\gamma}(\bbox{k_3}) G^{A}_{\delta}(\bbox{k_1}+\bbox{k_2}+\bbox{k_3}) 
+ G^{R}_{\gamma}(\bbox{k_3}) G^{R}_{\delta}(\bbox{k_1}+\bbox{k_2}+\bbox{k_3}) 
\right] \:. \nonumber
\end{eqnarray}
It is clear this expression could be obtained calculating the diagrams in $\bbox{k}$-space from the beginning.

Neglecting the rapidly-oscillating terms $G^{A}G^{A}$ and $G^{R}G^{R}$, we get
\begin{eqnarray}
\label{ds2kq}
\Delta\sigma_2 &=& - \frac{e^2 \hbar}{4\pi m^2} \sum_{\alpha\beta\gamma\delta}
W^2_{\alpha\beta\gamma\delta} \tau_\alpha \tau_\beta
\nonumber \\
&\times& \left\{
\sum\limits_{\bf{q} \bf{k} \bf{k'}} 
\bbox{k} \cdot (\bbox{k} - \bbox{q}) \:
\left[G^{A}_{\alpha}(\bbox{k}) G^{R}_{\beta}(\bbox{q} - \bbox{k}) +
G^{R}_{\alpha}(\bbox{k}) G^{A}_{\beta}(\bbox{q} - \bbox{k})\right]
G^{A}_{\gamma}(\bbox{k'}) G^{R}_{\delta}(\bbox{k'} + \bbox{q})
\right.
\\
&+& 
\left. 
\sum\limits_{\bf{q} \bf{k} \bf{k'}} 
\bbox{k} \cdot \bbox{k'}
\left[
G^{A}_{\alpha}(\bbox{k}) G^{R}_{\gamma}(\bbox{q} - \bbox{k}) 
G^{A}_{\beta}(\bbox{k'}) G^{R}_{\delta}(\bbox{q} - \bbox{k'})
+
G^{R}_{\alpha}(\bbox{k}) G^{A}_{\gamma}(\bbox{q} - \bbox{k}) 
G^{R}_{\beta}(\bbox{k'}) G^{A}_{\delta}(\bbox{q} - \bbox{k'})
\right] 
\right\}\:. \nonumber
\end{eqnarray}
Note that both sums over $\bf{q} \bf{k} \bf{k'}$ converge and, hence, can be calculated separately.

Taking into account that the sums over $\bbox{k}$ containing the products $G^{A}_{\alpha}(\bbox{k}) G^{R}_{\beta}(\bbox{q} - \bbox{k})$ are diagonal in subband indices ($\sim \delta_{\alpha\beta}$) due to the conditions~(\ref{goodconductivity},\ref{largedelta}), we obtain

\begin{equation}
\label{ds2q}
\Delta\sigma_2 = - \frac{e^2}{\pi \hbar} \int {d^2q \over (2\pi)^2} \sum_{\alpha\beta} {\tau_\alpha\tau_\beta \over \tau_{\alpha\beta}^2}
\left[ 
l_\alpha^2 \tilde{P}_\alpha(q) P_\beta(q) - l_\alpha l_\beta Q_\alpha(q) Q_\beta(q)
\right] \:.
\end{equation}
Here
\begin{eqnarray}
\label{Pq}
P_{\alpha}(q)&=&
{\hbar \over 2 \pi N_F \tau_\alpha} \sum_{\bf k} G^{A}_{\alpha}(\bbox{k})
G^{R}_{\alpha}(\bbox{k} - \bbox{q}) \:, \nonumber
\\ 
\tilde{P}_\alpha(q) &=&
{\hbar \over 2 \pi N_F \tau_\alpha} \sum_{\bf k} G^{A}_{\alpha}(\bbox{k})
G^{R}_{\alpha}(\bbox{k} - \bbox{q}) 
\left( 1 - {\bbox{q \cdot k} \over k_\alpha^2}\right)
\:, \\
Q_{\alpha}(q)&=&
{\hbar \over 2 \pi N_F \tau_\alpha} \sum_{\bf k} 
i \cos{(\varphi_{\bf k} - \varphi_{\bf q})} \:
G^{A}_{\alpha}(\bbox{k}) G^{R}_{\alpha}(\bbox{k} - \bbox{q}) \:,
\nonumber
\end{eqnarray}
and  $\varphi_{\bf k}$ is the angular coordinate of the vector ${\bbox k}$.

For the case of one occupied subband, the numerical calculation shows that only $q \ll k_{\rm F}$ give a contribution to the integral in~(\ref{ds2q}). In this range
\[
\tilde{P}(q) \approx P(q) \approx {1 \over \sqrt{1+(ql)^2}} \:,
\hspace{1cm}
Q(q) \approx {1 - P(q) \over ql} \:,
\]
and integration in~(\ref{ds2q}) yields
\begin{equation}
\label{ds2final}
\Delta\sigma_2 = - \frac{e^2}{\pi^2 \hbar} \ln{2}\:.
\end{equation}

\end{appendix}

\begin{figure}
\caption{
\label{f1}
The diagrams contributing to anomalous magnetoresistance, a) - d), and
similar to them, e).}
\end{figure}
\begin{figure}
\caption{
\label{f2}
The dependence of the conductivity correction, $\Delta\sigma$, on magnetic
field at various intersubband scattering rates and different level
occupations.}
\end{figure}
\begin{figure}
\caption{
\label{f3}
The dependence of the conductivity correction in zero field,
$\Delta\sigma(0)$, on the subband concentration ratio at different
intersubband scattering times, $\tau_1/\tau_{12}=0$ (solid curve),
$\tau_1/\tau_{12}=0.1$ (dashed curve), and $\tau_1/\tau_{12}=0.5$
(dotted one).}
\end{figure}
\begin{figure}
\caption{
\label{f4}
The dependence of the high-field limit asymptotics of $\Delta\sigma$
on the subband concentration ratio at different
intersubband scattering times, $\tau_1/\tau_{12}=0$ (solid curve),
$\tau_1/\tau_{12}=0.1$ (dashed curve), and $\tau_1/\tau_{12}=0.5$
(dotted one).}
\end{figure}

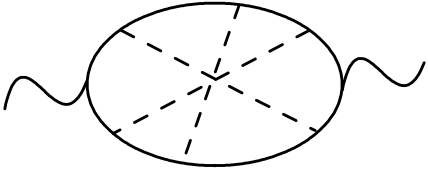
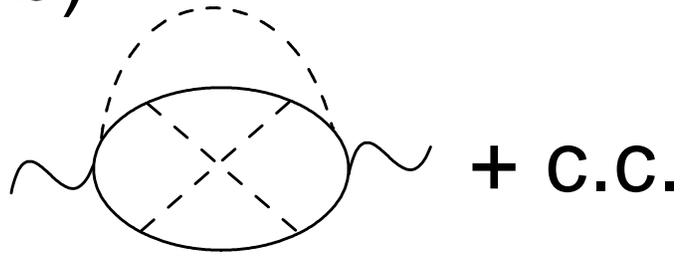
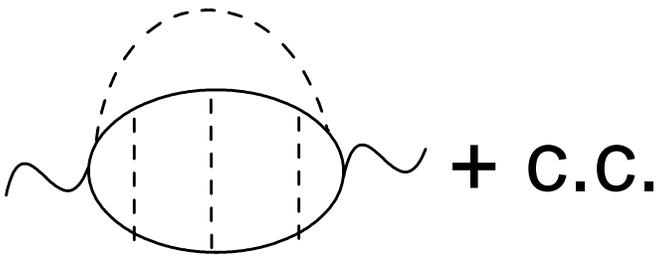
\end{document}